# Appeal and Scope of Misinformation Spread by AI Agents and Humans

*Emergent Research Forum (ERF) Paper*

**Lynnette Hui Xian Ng**
Carnegie Mellon University
lynnetteng@cmu.edu

**Wenqi Zhou**
Duquense University
zhouw@duq.edu

**Kathleen M. Carley**
Carnegie Mellon University
kathleen.carley@cs.cmu.edu

## Abstract

This work examines the influence of misinformation and the role of AI agents, called bots, on social network platforms. To quantify the impact of misinformation, it proposes two new metrics based on attributes of tweet engagement and user network position: *Appeal*, which measures the popularity of the tweet, and *Scope*, which measures the potential reach of the tweet. In addition, it analyzes 5.8 million misinformation tweets on the COVID-19 vaccine discourse over three time periods: Pre-Vaccine, Vaccine Launch, and Post-Vaccine. Results show that misinformation was more prevalent during the first two periods. Human-generated misinformation tweets tend to have higher appeal and scope compared to bot-generated ones. Tweedie regression analysis reveals that human-generated misinformation tweets were most concerning during Vaccine Launch week, whereas bot-generated misinformation reached its highest appeal and scope during the Pre-Vaccine period.

## Introduction

In recent years, misinformation has surged in numbers and influence during major public events, particularly during global public health crises like the COVID-19 pandemic. A 2020 study by the Pew Research Center found that nearly 48% of Americans encountered online misinformation about the COVID-19 virus. The uncertainty surrounding unprecedented events and public fears create fertile ground for misinformation to spread rapidly. However, misinformation is not a new phenomenon–it had existed long before the digital age through traditional media and word-of-mouth. A more concerning factor in the digital era is the role of AI agents, or Bots, in promoting misinformation at low cost and to a great extent. Unlike humans, bots can rapidly disseminate false content by exploiting platform algorithms that prioritize engagement and by leveraging social network structure to increase visibility and thus influence.

This study advances research on the influence of misinformation and the role of bots through two key approaches. First, it addresses the limitations of existing methods for quantifying the impact of misinformation, particularly that propagated through social networks. Previous studies relied on readily available engagement metrics from online platforms: favorites, retweets, and comments. To provide a more comprehensive assessment on the impact of misinformation at social networks, we integrated both message-level and network-level attributes to develop two novel metrics: appeal and scope, which capture the popularity and the reach of misinformation, two important aspects of influence. Using tweet data from X related to COVID-19 vaccine, we empirically demonstrate that these two new metrics more accurately quantify the influence of misinformation in engaging and reaching online social media audiences.

Second, we empirically examine the efficiency of bots and humans in spreading misinformation by addressing two main questions: a) How do bots and humans compete in spreading misinformation? and b) Whether and how do bots react to different time periods of critical public events?

We conducted descriptive analysis and applied the Tweedie regression model on COVID-19 tweets collected from three distinct time periods: Pre-Vaccine, Vaccine Launch, and Post-Vaccine. Our descriptive results reveal that the proportion of bots spreading misinformation increased over. Encouragingly, we found that regular tweets generally have higher appeal than their counterparts, misinformation tweets posted by the same account identity. Our Tweedie regression results reveal that the association between account identity (i.e., bots vs. humans) and misinformation influence (i.e., appeal and scope), which is shown to depend on the period of public events. We found opposite patterns of HumanMisinfo (misinformation tweets posted by humans) and BotsMisinfo (misinformation tweets posted by bots): BotMisinfo seemed to be more active and appealing during the Pre-Vaccine period, a time of heightened uncertainty, while HumanMisinfo reached its highest appeal and scope during the Vaccine Launch week, when vaccine rollout was released along with accurate and scientific updates. These results highlight the adaptive behavior of bots and the distinctive roles that bots and humans play in the dissemination of misinformation during critical public events.

## Context and Data

We adopted a published dataset that collected X conversations on the COVID-19 vaccine, termed here as CovidInfo (Blane et al, 2023). CovidInfo was divided into three time periods: a) Pre-Vaccine: the week before the vaccine rollout (12/01~12/07, 2020), 2) Vaccine Launch: the week of the release of the Pfizer vaccine (12/08~12/10, 2020), and 3) Post-Vaccine: a few weeks after the rollout (01/25~01/31, 2021).

Building on this dataset, we identified misinformation tweets by performing cosine similarity comparison of TwHIN-BERT-based vector embeddings of tweet content with corresponding vectors of expertly annotated misinformation tweets drawn from (Memon & Carley, 2020). Tweets that were classified as misinformation were those whose embeddings were at least 70% similar to the identified misinformation tweets, a threshold obtained from past work (Ng et. al., 2024). Categories of misinformation tweets that were annotated were: fake cure, conspiracy, fake treatment, false fact or prevention and false public health response. To differentiate the account identity between Bot and Human, we performed a bot-probability annotation using the BotHunter algorithm (Beskow and Carley, 2018), which used user-level and tweet-level features to predict the bot probability. The X account is considered as Bot if the probability is above 0.70, a threshold derived from past longitudinal analysis of stable bot scores (Ng et. al., 2022). We further derived CovidMisinfo by keeping only misinformation tweets from the CovidInfo dataset. The CovidMisinfo has a total of 5,890,967 misinformation tweets, of which 26.7% were posted in Pre-Vaccine, 41.1% during Vaccine Launch, and 32.2% Post-Vaccine.

We find that misinformation is more prevalent during the first two periods, accounting for a higher percentage of the total tweets (69.9% and 70.7% vs 63.4%). Interestingly, throughout all three periods, bots performed more diligently than humans in spreading misinformation, consistently posting a higher proportion of misleading content relative to their total tweet volume. At the account level, more human users engaged in the conversation about Covid-19 Vaccine during the Vaccine Launch week. As the vaccine rollout progressed, human user involvement decreased. However, bot accounts exhibited a more consistent pattern of engagement, with an increasing number of bots joining the discussion over time. As a result, the bot-to-human account ratio rose significantly during the Post-Vaccine period, from previously 35% to 38.2%, suggesting a potential effort to shape public opinion on the vaccine.

We follow AI agent literature to design our two key independent variables: whether the user $u$ was a bot ($IsBot_u$) and the posted period of the tweet $i$ ($TimePeriod_i$). We differentiate among the three time periods: Pre-Vaccine, Vaccine Launch, and Post-Vaccine. We also introduce several control variables. $MisinfoType_i$ represented the type of misinformation of the tweet $i$, which is shown to affect engagement of the tweet, and therefore virality. $IsRetweet_i$ indicated whether the tweet was a retweet or an original tweet, since retweet is an important feature that social media platforms use for content recommendation. Finally, since more established accounts are more influential, $AccountAge_u$ was included as the difference between the creation date of tweet $i$'s account $u$ and the last date of the $TimePeriod_i$.

## Appeal and Scope of Online Misinformation

We designed two metrics to capture the influence of misinformation spread through online social networks: *Appeal* and *Scope*. First, we constructed an all-communication network graph for each period $t$. Users that tweeted during $t$ were represented as nodes in the graph. Users were linked together if they had an interaction through a tweet, i.e., a retweet or a mention within the tweet. We calculated the *Total Degree Centrality* of each user $u$ at $t$, which is the sum of the number of incoming and outgoing interactions the user has. Second, we measured *Retweet Count* as the number of retweets.

Appeal captured the popularity of the tweet. This metric weights the tweet engagement by the relative influence of tweet $i$'s account owner at X. $Appeal_i = RetweetCount_i \times (1 + TotalDegreePercentile_{\{u,t\}})$, where $RetweetCount_i$ is the retweet count of tweet $i$, and $TotalDegreePercentile_{\{u,t\}}$ is the percentile of user $u$ based on total degree centrality at $t$. Scope represents the number of users that the tweet can potentially reach. This metric weights the network reach of the user $u$ by the spread of the tweet $i$. $Scope_i = TotalDegreeCentrality_{\{u,t\}} \times (1 + RetweetCountPercentile_{\{i,t\}})$, where $TotalDegreeCentrality_{\{u,t\}}$ is the total degree centrality of user $u$ at $t$, and $RetweetCountPercentile_{\{i,t\}}$ is the percentile of tweet $i$ based on the number of retweets it had at $t$.

To demonstrate these two proposed metrics and their difference, we utilized an all-communication network graph of the Pre-Vaccine period illustrated in Figure 1. In the network illustrated, green nodes represent bot accounts and blue nodes represent human accounts. The nodes are sized by the average retweet count of the tweets authored by the account; the larger the node, the more retweets a user's tweets have. The number of links between the nodes help illustrate total degree centrality. For example, two accounts, @*thebias_news* and @*GeodanNew*, are linked to a larger number of other users, meaning that their tweets can reach a larger group of audience. Therefore, they have higher scope than another two accounts, @*spectatorindex* and @*jaketapper*. In contrast, @*spectatorindex* and @*jaketapper*'s tweets have higher appeal, because although they do not have a wide set of interactions with a big group of others, their tweets have a lot of engagement indicated by larger node sizes.

Traditional engagement metrics, such as retweet counts, can be misleading when assessing a tweet's influence. For instance, while users like @*jaketapper* and @*spectatorindex* rank high based on their large node size, their tweets do not necessarily spread widely throughout the network. On the other hand, users like @*GeodanNew* and @*thebias_news* appear to have limited impact due to lower retweet counts, yet their tweets can disseminate extensively through an interactive network of engaged users built over time.

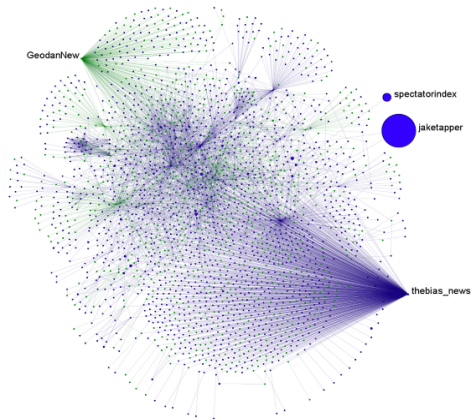

**Figure 1: Network Graph Snapshot during the Pre-Vaccine Period**

Figure 2 illustrates the average appeal and average scope with the natural log transformation among BotMisinfo, HumanMisinfo, BotInfo (regular tweets posted by bots), and HumanInfo tweets (regular tweets posted by humans). Misinformation tweets generally had lower appeal and scope compared to its counterpart–regular tweets posted by the same type of account (i.e., humans or bots)–with one notable exception: BotMisinfo tweets were nearly 34% more widespread than BotInfo tweets. In addition, humans surpassed bots in both appeal and scope across misinformation and regular tweets. However, this discrepancy was less pronounced for misinformation. HumanMisinfo had 22 times higher appeal than BotMisinfo, whereas HumanInfo had 449 times higher appeal than BotInfo, indicating the bots' disproportionately better performance in misinformation than the regular tweets.

We constructed the following two models to empirically evaluate the effectiveness of online misinformation messages posted by Bots and Humans on the social media:

Baseline Model: $Metric_i = \alpha_0 + \alpha_1 * IsBot_u + \alpha_2 * TimePeriod_i + \alpha_3 * MisinfoType_i + \alpha_4 * IsRetweet_i + \alpha_5 * AccountAge_u + \varepsilon_i$

Conditional Effect model: $Metric_i = \beta_0 + \beta_1 * IsBot_u + \beta_2 * TimePeriod_i + \beta_3 * IsBot_u * TimePeriod_i + \beta_4 * MisinfoType_i + \beta_5 * IsRetweet_i + \beta_6 * AccountAge_u + \delta_i$

,where $i$ denotes an individual tweet, $u$ denotes the account owner who posted the tweet $i$, and $Metric_i$ includes two DVs, $Appeal_i$ and $Scope_i$. Given the large number of zeros in both DVs, we estimated these two models for each DV using Tweedie Regression model with a parameter of 1.5, implying a compound Poisson-Gamma distribution for the DV. Estimation results of key variables are summarized in Table 1. The variance inflation factor scores for all variables in each model for each DV returned were below 5.

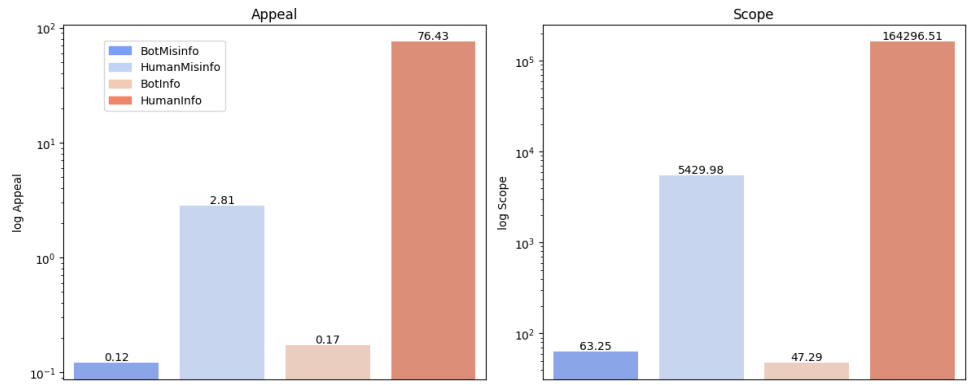

*Note: the original metric value, without log-transformation, is listed at the top of each bar.*
**Figure 2: Average Appeal and Scope of Tweets**

|  | Baseline Model | | Conditional Effect Model | |
|---|---|---|---|---|
|  | Appeal | Scope | Appeal | Scope |
| Bot=1 | -2.42 *** | -2.26 *** | -2.19 *** | -1.64 *** |
| Vaccine Launch=1 | 6.70E-2 *** | 0.42*** | 0.13*** | 0.62*** |
| Post-Vaccine=1 | 4.38E-2** | 0.23*** | 9.59E-2*** | 0.38*** |
| Bot * Vaccine Launch |  |  | -0.36*** | -0.98*** |
| Bot * Post-Vaccine |  |  | -0.28*** | -0.68*** |
| *Notes: *** for p<0.001, ** for p<0.01, *for p<0.05.* | | | | |

**Table 1. Tweedie Regression Results from the CovidMisinfo data set**

The results from the Baseline Model indicates that, on average, misinformation tweets were more concerning during the Vaccine Launch week. During this period, misinformation tweets exhibit 6.92% higher appeal and 52.04% greater scope compared to other periods. Overall, humans outperformed bots in spreading misinformation and engaging audiences. Specifically, BotMisinfo showed significantly lower appeal and scope than HumanMisinfo by 91.11% and 89.56%, respectively.

Results from the Conditional Effect Model provide a more granular view. Being consistent with the Baseline Model's, HumanMisinfo was indeed more widespread and popular during the Vaccine Launch week compared to other two periods, with appeal increased by 14.11% and scope by 85.52%, indicated by the coefficients on Vaccine Launch. In contrast, BotMisinfo exhibits an opposite pattern. The coefficients on two interaction terms indicate that BotMisinfo was least popular and widespread during the Vaccine Launch week, with Pre-Vaccine period being the most influential time period. This finding complements the results from Baseline Model, highlighting the importance of studying the temporal conditional effect. When comparing misinformation spread by humans and bots, regardless of time period, HumanMisinfo consistently outperformed BotMisinfo in both appeal and scope. Their difference is most pronounced during the Vaccine Launch week, during which BotMsinfo was 30.2% lower in appeal, and 62.3% lower in scope. This is echoed with our earlier descriptive observation that humans were more actively involved in posting misinformation tweets during the Vaccine Launch week, as reflected in the higher volume of HumanMisinfo relative to BotMisinfo.

## Discussion

This work proposes two robust metrics, appeal and scope, for measuring the influence of online misinformation delivered through social networks. Grounded in literature of online user-generated contents and AI, we demonstrated the unique advantage of these two metrics using a large COVID-vaccine tweet dataset. Compared to regular tweets from the same type of account (humans or bots), misinformation tweets generally had lower appeal and scope. However, a notable exception is that BotMisinfo tweets were nearly 34% more widespread than BotInfo tweets. Regression analyses focusing on misinformation found that bots were generally less influential than humans in terms of both appeal and scope of their generated misinformation. HumanMisinfo had the highest appeal and scope during Vaccine Launch week, whereas BotsMisinfo had the lowest compared to Pre- and Post-Vaccine periods. Extensions of this work included generalizing the methodology to other events and agent types.

## Acknowledgements

The first and third authors are supported by the Scalable Technologies for Social Cybersecurity, U.S. Army (W911NF20D0002), the Minerva-Multi-Level Models of Covert Online Information Campaigns (N000142112765), Threat Assessment Techniques for Digital Data (N000142412414), and MURI (N000142112749), Office of Naval Research.